\documentclass[conference]{IEEEtran}  
\usepackage[dvips]{graphicx}
\usepackage{epsfig,amssymb,amsmath,color}
\newcommand{\beq}{\begin{equation}}
\newcommand{\eeq}{\end{equation}}
\newcommand{\beqn}{\begin{eqnarray}}

\newcommand{\eeqn}{\end{eqnarray}}
\def\bmath#1{\mbox{\boldmath$#1$}}

\long\def\symbolfootnote[#1]#2{\begingroup%
\def\thefootnote{\fnsymbol{footnote}}\footnote[#1]{#2}\endgroup}

\title{Fundamental Limitations of Pixel Based Image Deconvolution in Radio Astronomy}
\author{\authorblockN{Sarod Yatawatta$^{1,2,3}$}
}
\begin{document}
\maketitle
\symbolfootnote[0]{$^1$Kapteyn Astronomical Institute, University of Groningen, Groningen, The Netherlands}
\symbolfootnote[0]{$^2$ASTRON, Dwingeloo, The Netherlands}
\symbolfootnote[0]{$^3$yatawatta@astron.nl}
%

\begin{abstract}
Deconvolution is essential for radio interferometric imaging to produce scientific quality data because of finite sampling in the Fourier plane. Most deconvolution algorithms are based on CLEAN which uses a grid of image pixels, or clean components. A critical matter in this process is the selection of pixel size for optimal results in deconvolution.  As a rule of thumb, the pixel size is chosen smaller than the resolution dictated by the interferometer. For images consisting of unresolved (or point like) sources, this approach yields optimal results. However, for sources that are not point like, in particular for partially resolved sources, the selection of right pixel size is still an open issue. In this paper, we investigate the limitations of pixelization in deconvolving extended sources. In particular, we pursue the usage of orthonormal basis functions to model extended sources yielding better results than by using clean components.
\end{abstract}
\begin{keywords}
Radio interferometry, Image deconvolution, Estimation theory.
\end{keywords}

\section{Introduction}
Due to the incomplete Fourier plane sampling of radio synthesis observations, deconvolution is essential for making high fidelity images. The traditional CLEAN \cite{Hogbom} algorithm and its variants are widely used for this deconvolution. There are several limitations of clean being applied to a typical image. First, the centroids of point sources will not exactly match pixel coordinates on a regular grid. Secondly, some sources might be extended, thus requiring more than one clean component.

Therefore, since its invention, several studies on the performance of clean and its limitations have been conducted. In \cite{Schwarz} for instance, the convergence and residual errors in terms of a least squares fit for the Fourier plane data is discussed. The work  \cite{Briggs} (ch. 6) focuses mainly on clean components and their analogy to Fourier components of the sky image. The fundamental limitations of image pixelization (or having a regular grid of clean components) is studied in \cite{PIX}, especially in the case where sources are located off a pixel center.

In this paper, we focus on improving the deconvolution of bright, extended sources that are barely resolved. In order to arrive at our results, we use statistical estimation theory to derive some fundamental limits of clean in deconvolving such sources. Compared to the work of \cite{PIX} which give numerical bounds, we derive analytical bounds on its performance. Moreover, compared to \cite{Briggs} which takes a deterministic approach to study clean component placement, we take a statistical approach to derive the Cramer-Rao lower bound \cite{Kay,Behery}. 

The limitations of deconvolving an extended source with a set of clean components have been overcome by using clean components that have different scale sizes. For instance, \cite{MSCLEAN},\cite{Leshem10} gives a comprehensive overview of this approach and comparisons with similar existing approaches. However, using multi scale pixels would still be limited by the resolution limit of the interferometer in case of  barely resolved sources. In order to overcome this deficiency, we consider using a two dimensional orthonormal basis instead of a set of clean components. As a real example of this technique, we select one such basis called the shapelet basis \cite{SHP1} and apply this technique to Westerbork Synthesis Radio Telescope (WSRT) low frequency observations. Shapelets have  been extensively used in astronomical image processing applications \cite{SHP1},\cite{SHP4}, including deconvolution of radio interferometric data \cite{SHP3}. In this paper, we use shapelets for high fidelity imaging when clean based algorithms perform poorly.
\section{Mathematical preliminaries}
In this section we derive some fundamental limitations of clean and try to explain the reason that an orthonormal basis could improve on using clean components to model extended structure. For simplicity, we first consider a one dimensional image and its corresponding Fourier plane (axis). The image axis is given by $l$ and the corresponding visibility axis is $u$. The corresponding units are radians and wavelengths, respectively.
\subsection{Interferometric imaging}
Let us consider observing a point source at the origin, whose brightness is given by $\delta(l)$ (the Dirac delta function). The visibilities correspond to the Fourier transform of $\delta(l)$, which is $1$. We only observe at a set of discrete points on the $u$ axis, and this is equal to sampling by the weighted sampling function
\beq
\Pi(u)=\sum_i w(u_i) \delta(u-u_i)
\eeq
where $w(u_i)$ correspond to the weight we assign to the $i$-th sampling point, which is at $u_i$ on the $u$ axis. For the remainder we consider all weights to be unity. The observed image $I(l)$ is the inverse discrete Fourier transform
\beqn
I(l)&=&\sum_k \biggl(\sum_i 1 \delta(u-u_i) \biggr) \exp(j2\pi l u_k)\\\nonumber
 &=& \sum_k  \exp(j2\pi l u_k)
\eeqn
which is the point spread function (PSF). In order to denote the nominal resolution limit, we use $b=1/max(|u|)$.
\subsection{Pixelization error}
Now, let us consider a point source which is displaced by $l_0$ from the origin, which has $\gamma_0$ brightness, which is represented by $\gamma_0 \delta(l-l_0)$. The sampled visibility at the $i$-th point on the $u$ axis is given by
\beq \label{vis}
y_i=\gamma_0 \exp(-j2\pi l_0 u_i) + n_i
\eeq
where $n_i$ is the observation noise. We assume the noise to be white, uncorrelated complex (circular) Gaussian with zero mean and variance $\sigma^2$.

Due to pixelization, we represent this point source with the pixel at the origin, if $l_0$ is small enough. We estimate the magnitude $\alpha$ of the clean component at the origin by minimizing the least squared error. The error at the $i$-th sampling point will be 
\beq
\xi_i=\gamma_0 \exp(-j2\pi l_0 u_i) + n_i - \alpha
\eeq

and the total (average) error to be minimized is
\beq \label{E}
\xi^2=\frac{1}{N}\sum_i E\{ \xi_i \xi^{\star}_i \}=\frac{1}{N} \sum_i \gamma_0^2+\sigma^2+\alpha^2-2 \alpha\gamma_0 \cos(2\pi l_0 u_i)
\eeq
where $N$ is the total number of sampling points on the $u$ axis.

The solution for $\alpha$ is obtained by solving $\frac{\partial \xi^2}{\partial \alpha}=0$
\beq \label{ahat}
\hat{\alpha}=\frac{\gamma_0}{N} \sum_i \cos(2\pi l_0 u_i)
\eeq
and substituting  (\ref{ahat}) to (\ref{E}) gives the minimum error, $\hat{\xi}_i = \gamma_0 \exp(-j2\pi l_0 u_i) + n_i - \hat{\alpha}$.
This error can be minimized by shifting the pixel grid by $l_0$, as shown in \cite{PIX}. Hence, this does not cause a real problem in deconvolution.

\subsection{Clean component placement at arbitrary locations}
We relax the pixelization requirement and assume we could place a clean component at any location. As before, we observe a point source, with magnitude $\gamma_0$, positioned at $l_0$ on the image axis. The noisy observed visibilities are given by (\ref{vis}).

Let ${\bf y}\buildrel\triangle\over=[y_1,y_2,\ldots,y_N]^T$ and ${\bf u}\buildrel\triangle\over=[u_1,u_2,\ldots,u_N]^T$ represent the vectorized forms of the observed visibilities and the observation coordinates, respectively. We assume there are $N$ observation points, the aforementioned vectors have dimensions $N\times1$. The likelihood of the observation is given by
\beqn \label{probl}
\lefteqn{p({\bf y}|\sigma^2,l_0,\gamma_0,{\bf u})}&&\\\nonumber
&\mbox{}=&\frac{1}{(\pi \sigma^2)^N} \exp\biggl(\frac{-1}{\sigma^2} \sum_i |y_i-\gamma_0\exp(-j 2 \pi l_0 u_i)|^2\biggr)
\eeqn

 The variance of estimating $l_0$ and $\gamma_0$  (or the Cramer-Rao lower bound) are given by
\beq \label{Varl}
Var\bigl( {\hat l_0} \bigr)\geq  \frac{\sigma^2}{8 \gamma_0^2 \pi^2 \sum_i u_i^2},\ \ Var\bigl({\hat \gamma_0}\bigr) \geq \frac{\sigma^2 }{2 N}
\eeq
Note that similar results have been derived for a single interferometer \cite{Behery}, or a single point in ${\bf u}$.  We see from (\ref{Varl}) that the error in estimation of $l_0$ is not only dependent on the noise $\sigma^2$, but also dependent on the sampling points on the ${\bf u}$ axis, which is the resolution limit of the interferometer.

\subsection{Partially resolved sources}
\begin{figure}[htbp]
\begin{minipage}{0.98\linewidth}
\begin{minipage}{0.48\linewidth}
\centering
 \centerline{\epsfig{figure=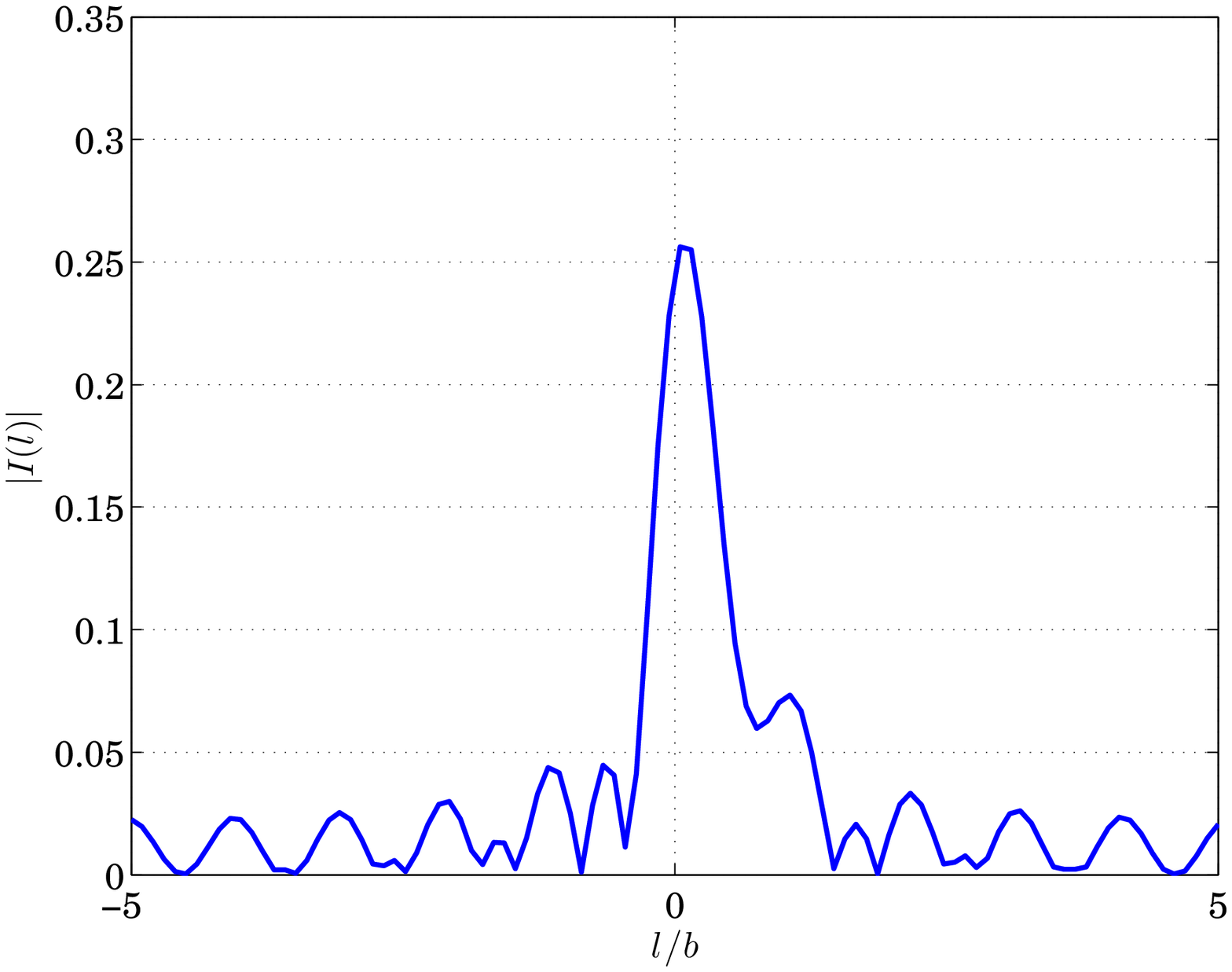,width=3.7cm}}
\vspace{0.1cm} \centerline{(a)}\smallskip
\end{minipage}
\begin{minipage}{0.48\linewidth}
\centering
 \centerline{\epsfig{figure=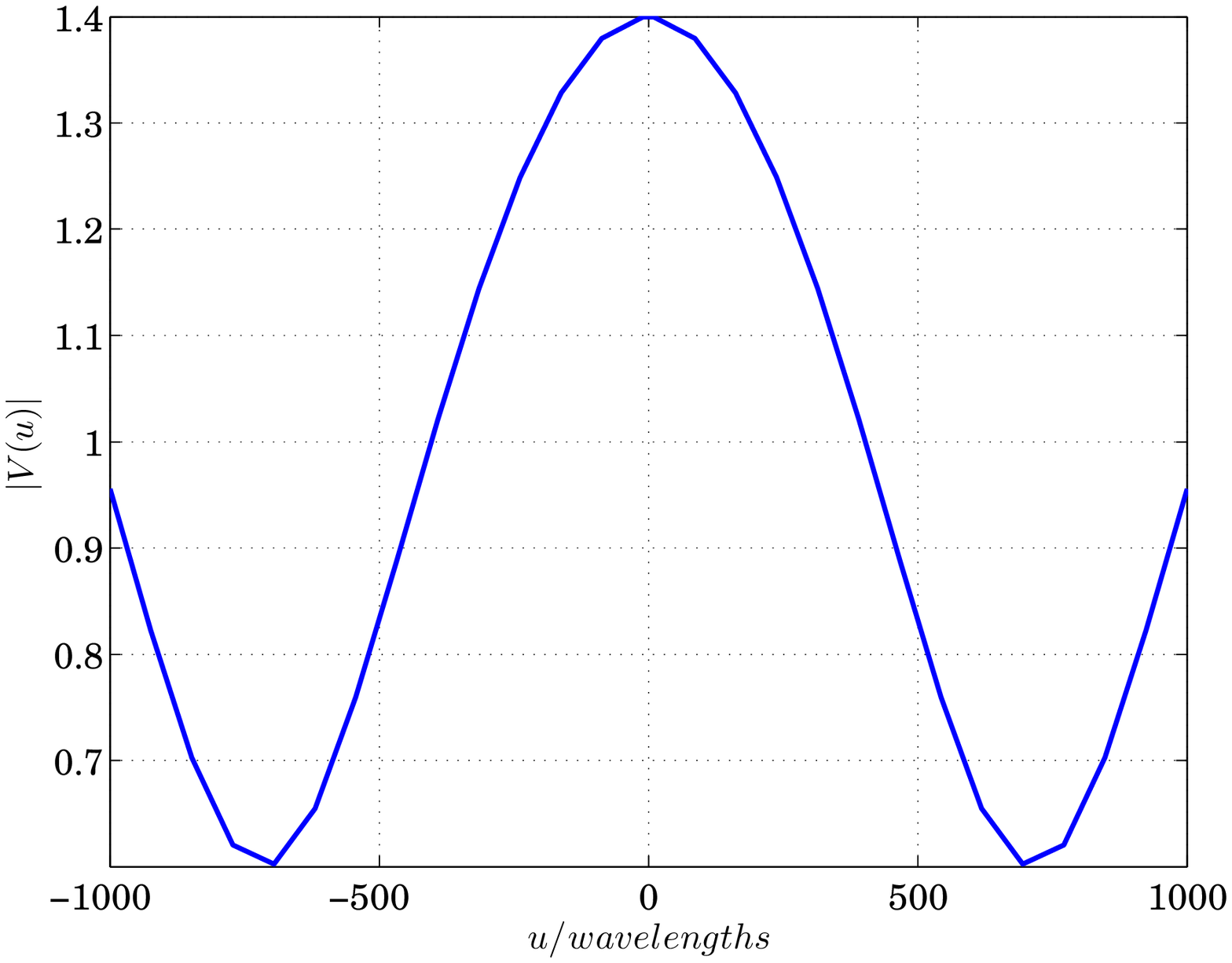,width=3.7cm}}
\vspace{0.1cm} \centerline{(b)}\smallskip
\end{minipage}
\end{minipage}
\caption{Two point sources, almost unresolved case. The image (a) and the corresponding visibility amplitudes (b) are given. The magnitudes are $\gamma_0=1.0$, $\gamma_1=0.4$. The positions are $l_0=0.1b$, $l_1=0.8b$, where $b$ is the resolution. Peaks are clearly seen at positions close to $l_0$ and $l_1$.\label{two_resolved}}
\end{figure}

\begin{figure}[htbp]
\begin{minipage}{0.98\linewidth}
\begin{minipage}{0.48\linewidth}
\centering
 \centerline{\epsfig{figure=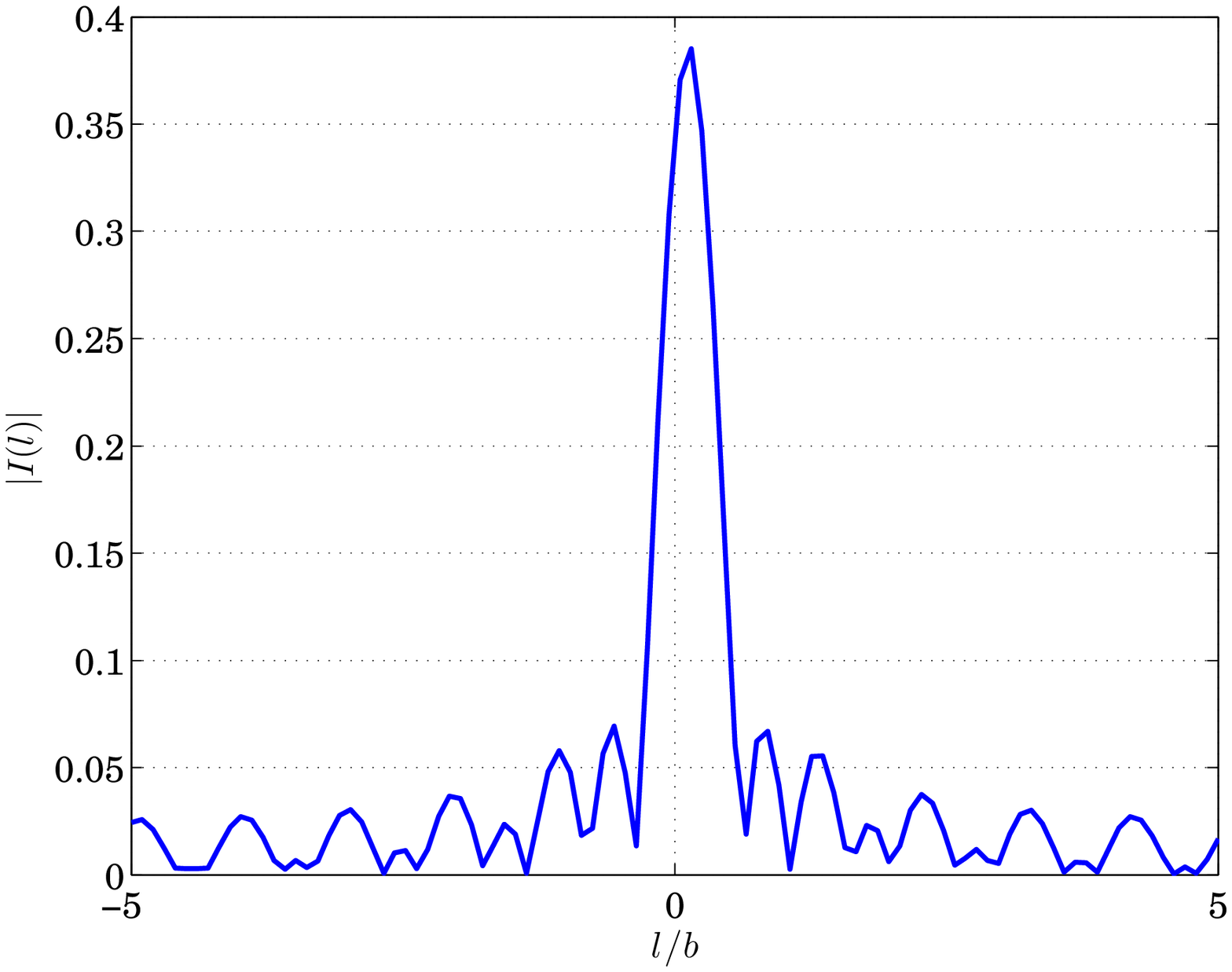,width=3.7cm}}
\vspace{0.1cm} \centerline{(a)}\smallskip
\end{minipage}
\begin{minipage}{0.48\linewidth}
\centering
 \centerline{\epsfig{figure=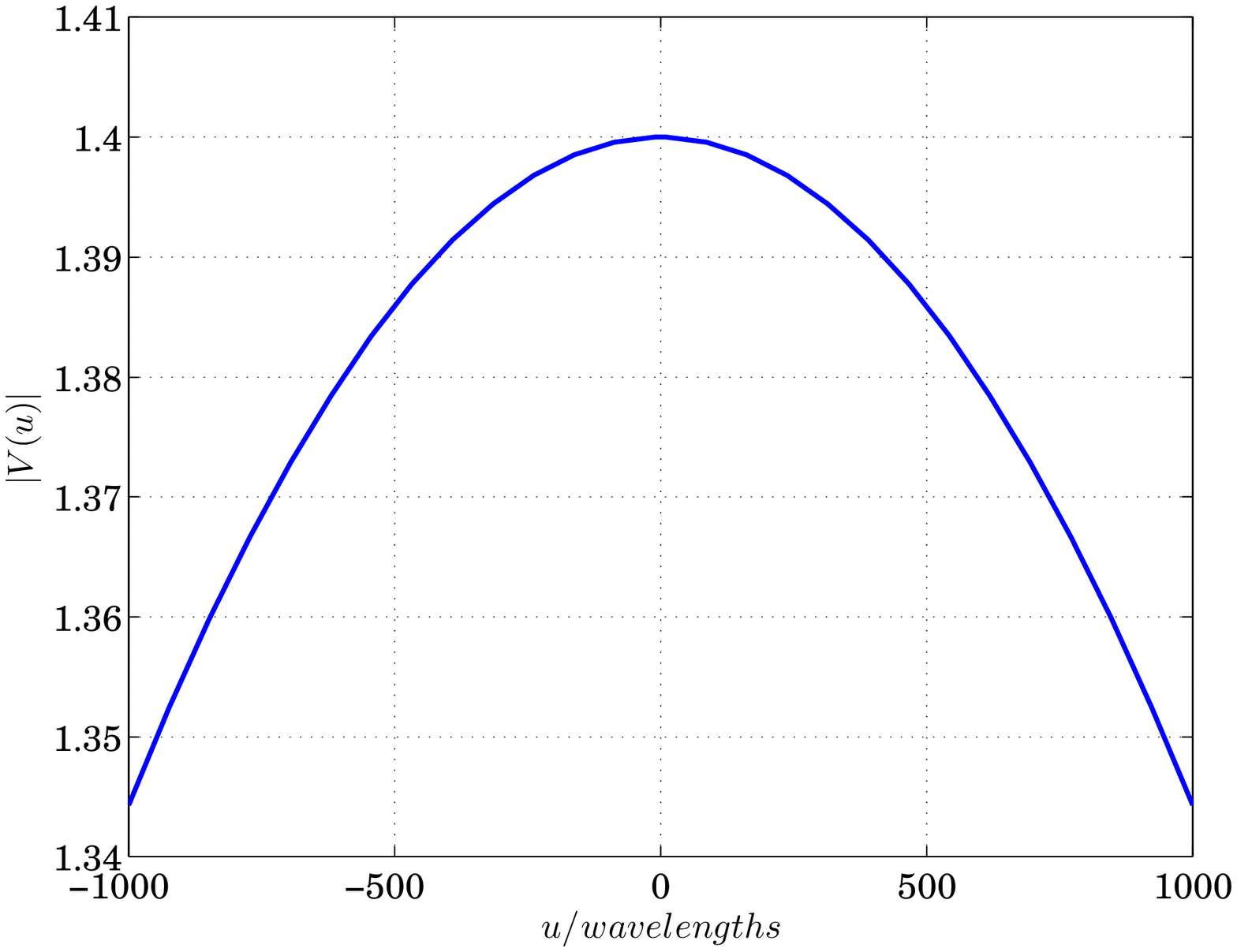,width=3.7cm}}
\vspace{0.1cm} \centerline{(b)}\smallskip
\end{minipage}
\end{minipage}
\caption{Two point sources, unresolved case. The image (a) and the corresponding visibility amplitudes (b) are given. The magnitudes are $\gamma_0=1.0$, $\gamma_1=0.4$. The positions $l_0=0.1b$, $l_1=0.2b$ are too close to be resolved due to the finite resolution $b$.\label{two_unresolved}}
\end{figure}

The more challenging case in deconvolution is when the source cannot be represented as a pure point source, or as a single clean component. The simplest example for this is having two sources, with magnitudes $\gamma_0$ and $\gamma_1$, shifted by $l_0$ and $l_1$, respectively. The observed (dirty) image is 
\beq
I(l)=\sum_i \biggl( \gamma_0 \exp(-j2\pi l_0 u_i)+ \gamma_1 \exp(-j2\pi l_1 u_i) \biggr) \exp(j2\pi l u_i) 
\eeq
The ability to correctly estimate the magnitudes and positions is dependent on the sampling on the $u$ axis. For some cases, we will be unable to estimate them accurately, as we shall see later. Obviously, this happens when the two sources become closer together. In Figs. \ref{two_resolved} and \ref{two_unresolved}, we have presented dirty images (and the corresponding visibility amplitudes) for barely resolved and unresolved cases, respectively.

As usual, the observed visibilities are given by
\beq \label{vis2}
y_i=\gamma_0 \exp(-j2\pi l_0 u_i)+ \gamma_1 \exp(-j2\pi l_1 u_i) + n_i
\eeq
and the likelihood is
\beqn \label{probl2}
\lefteqn{p({\bf y}|\sigma^2,l_0,l_1,\gamma_0,\gamma_1,{\bf u})}&&\\\nonumber
&=&\frac{1}{(\pi \sigma^2)^N} \exp\biggl(\frac{-1}{\sigma^2} \sum_i |y_i-\gamma_0\exp(-j 2 \pi l_0 u_i)\\\nonumber
&&- \gamma_1\exp(-j 2 \pi l_1 u_i)|^2\biggr)
\eeqn

\begin{figure}[htbp]
\begin{minipage}{0.98\linewidth}
\begin{minipage}{0.48\linewidth}
\centering
 \centerline{\epsfig{figure=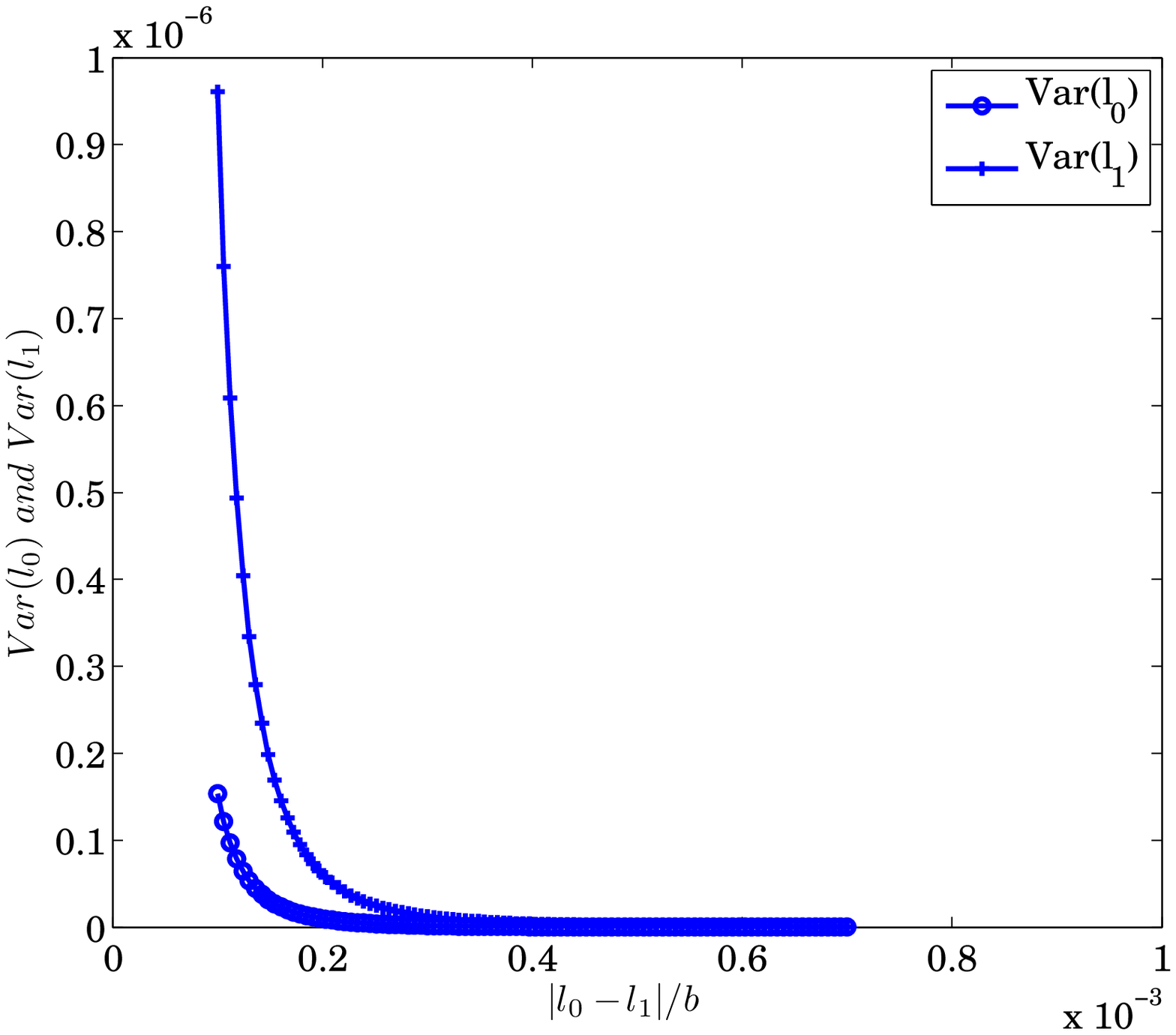,width=3.7cm}}
\vspace{0.1cm} \centerline{(a)}\smallskip
\end{minipage}
\begin{minipage}{0.48\linewidth}
\centering
 \centerline{\epsfig{figure=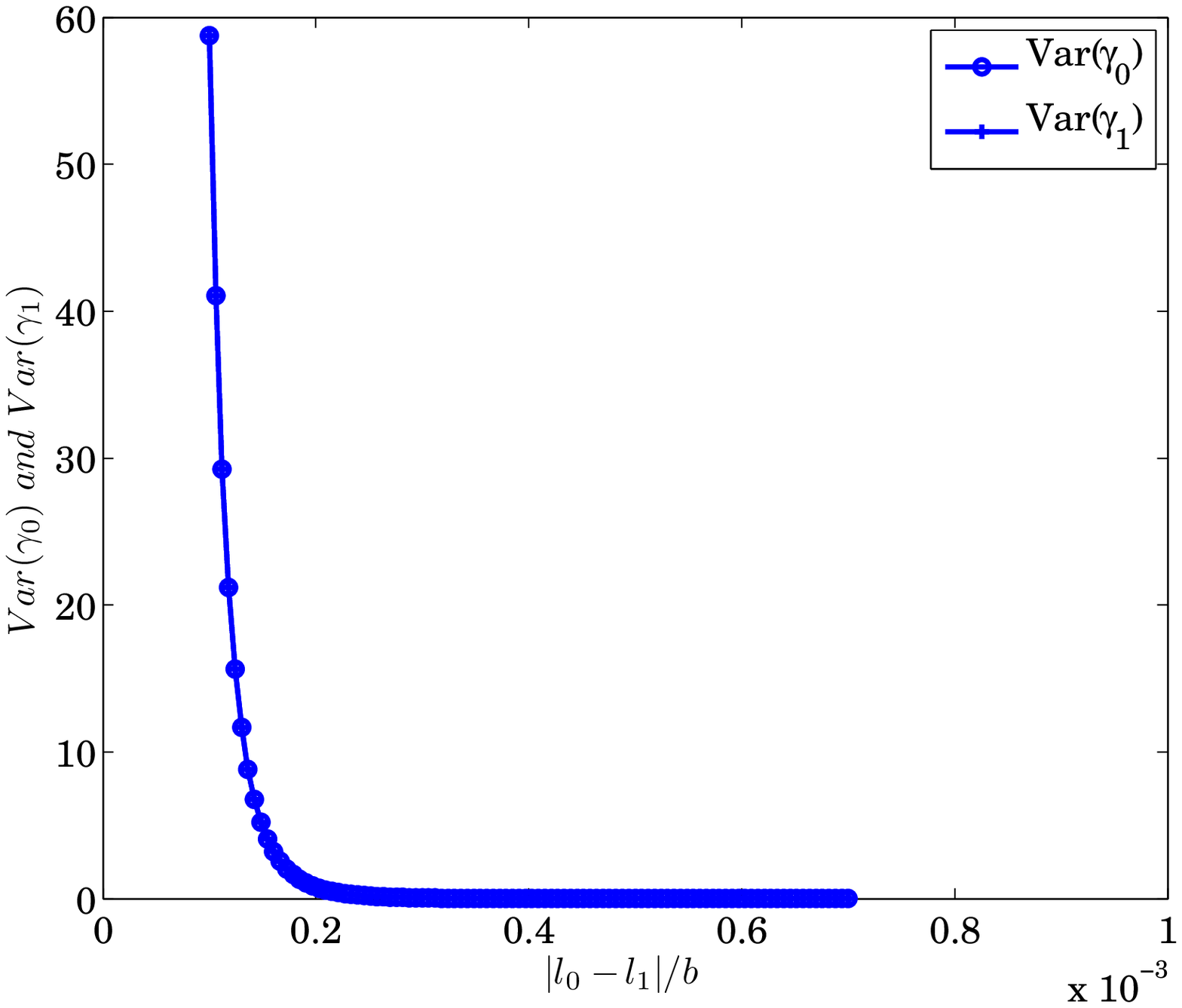,width=3.7cm}}
\vspace{0.1cm} \centerline{(b)}\smallskip
\end{minipage}
\end{minipage}
\caption{Variation of the Cramer Rao lower bound with the spacing between two point sources. The variance in estimating the positions $l_0$,$l_1$ are given in (a) and the variance in estimating the magnitudes $\gamma_0$ and $\gamma_1$ are given in (b). The true magnitudes are $\gamma_0=1.0$, $\gamma_1=0.4$ while the noise variance is $\sigma^2=0.2^2$. It is clearly seen that as the two sources come closer than about $0.4b$, the variance (or the estimation error) increases significantly.\label{two_crlb}}
\end{figure}

The ML estimate is obtained by minimizing the cost $J$
\beqn \label{Jcost2}\nonumber
\lefteqn{J}&=&\frac{-1}{\sigma^2}\sum_i |y_i|^2-y_i\biggl(\gamma_0 \exp(j2\pi l_0 u_i)+ \gamma_1 \exp(j2\pi l_1 u_i)\biggr)\\\nonumber
&&-y_i^{\star}\biggl(\gamma_0 \exp(-j2\pi l_0 u_i) + \gamma_1 \exp(-j2\pi l_1 u_i)\biggr) \\ &&+\gamma_0^2+\gamma_1^2+2 \gamma_0\gamma_1  \cos\bigl(2\pi (l_0-l_1) u_0 \bigr)
\eeqn
with respect to $l_0$,$l_1$,$\gamma_0$, and $\gamma_1$. Let ${\bmath \theta}=[l_0,l_1,\gamma_0,\gamma_1]^{T}$ be the parameter vector to be estimated. Then the Fisher information matrix is given by
\beq \label{F2}
{\mathcal F}({\bmath \theta}) =-E\{ \frac{\partial}{\partial {\bmath \theta}}\frac{\partial}{\partial {\bmath \theta}^{T}} J \}
\eeq
and the Cramer Rao bound is given by the diagonal entries of the inverse of  ${\mathcal F}({\bmath \theta})$:
\beqn \label{crlb2}
Var(\hat{l}_0) \ge  [{\mathcal F}^{-1}({\bmath \theta})]_{1,1},\ \ Var(\hat{l}_1) \ge  [{\mathcal F}^{-1}({\bmath \theta})]_{2,2},\\\nonumber
Var(\hat{\gamma}_0) \ge  [{\mathcal F}^{-1}({\bmath \theta})]_{3,3},\ \ Var(\hat{\gamma}_1) \ge  [{\mathcal F}^{-1}({\bmath \theta})]_{4,4}. 
\eeqn
In Fig. \ref{two_crlb}, we have given the CRLB for our example case.

\begin{figure}[htbp]
\begin{minipage}{0.98\linewidth}
\begin{minipage}{0.48\linewidth}
\centering
 \centerline{\epsfig{figure=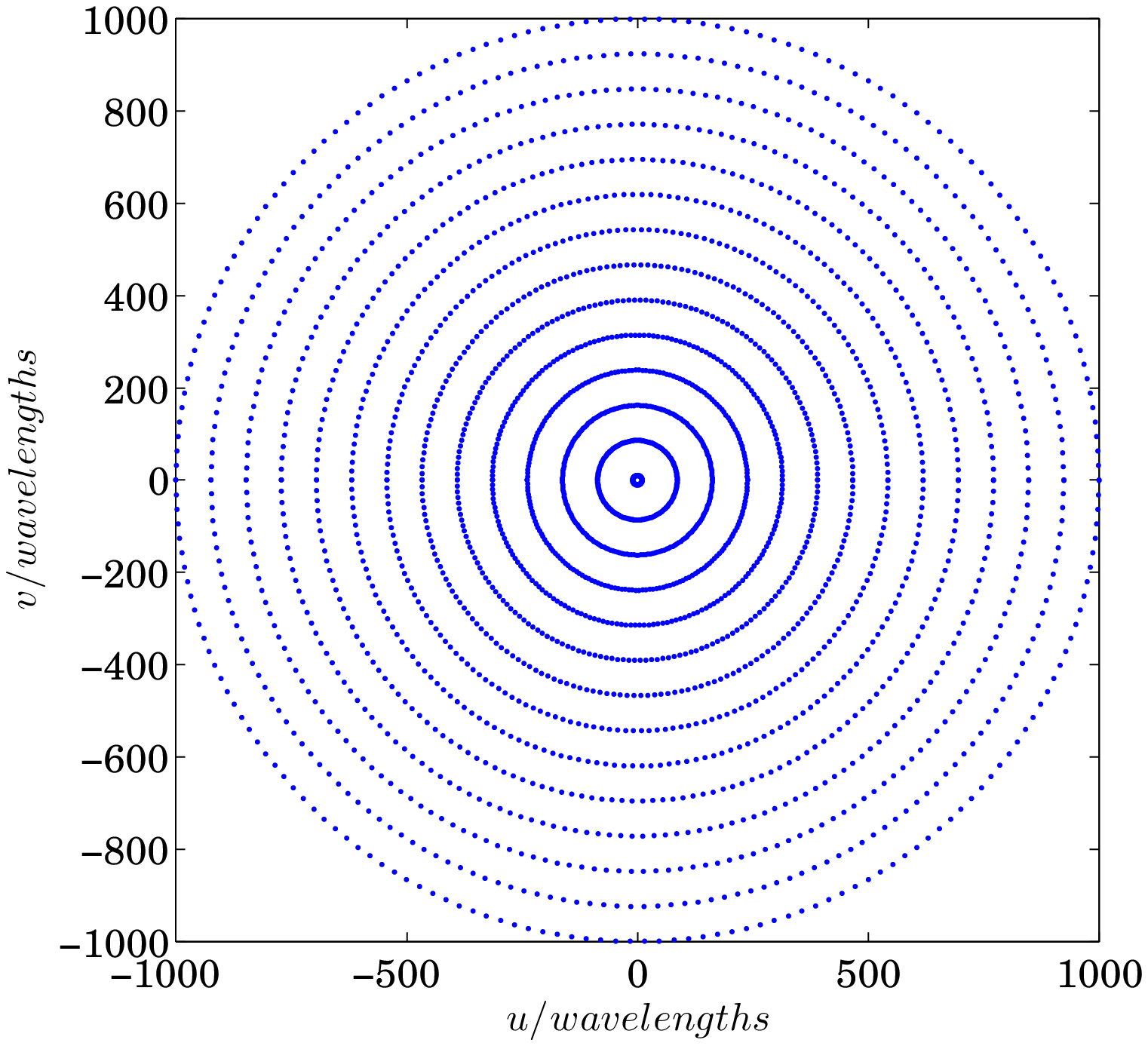,width=3.7cm}}
\vspace{0.1cm} \centerline{(a)}\smallskip
\end{minipage}
\begin{minipage}{0.48\linewidth}
\centering
 \centerline{\epsfig{figure=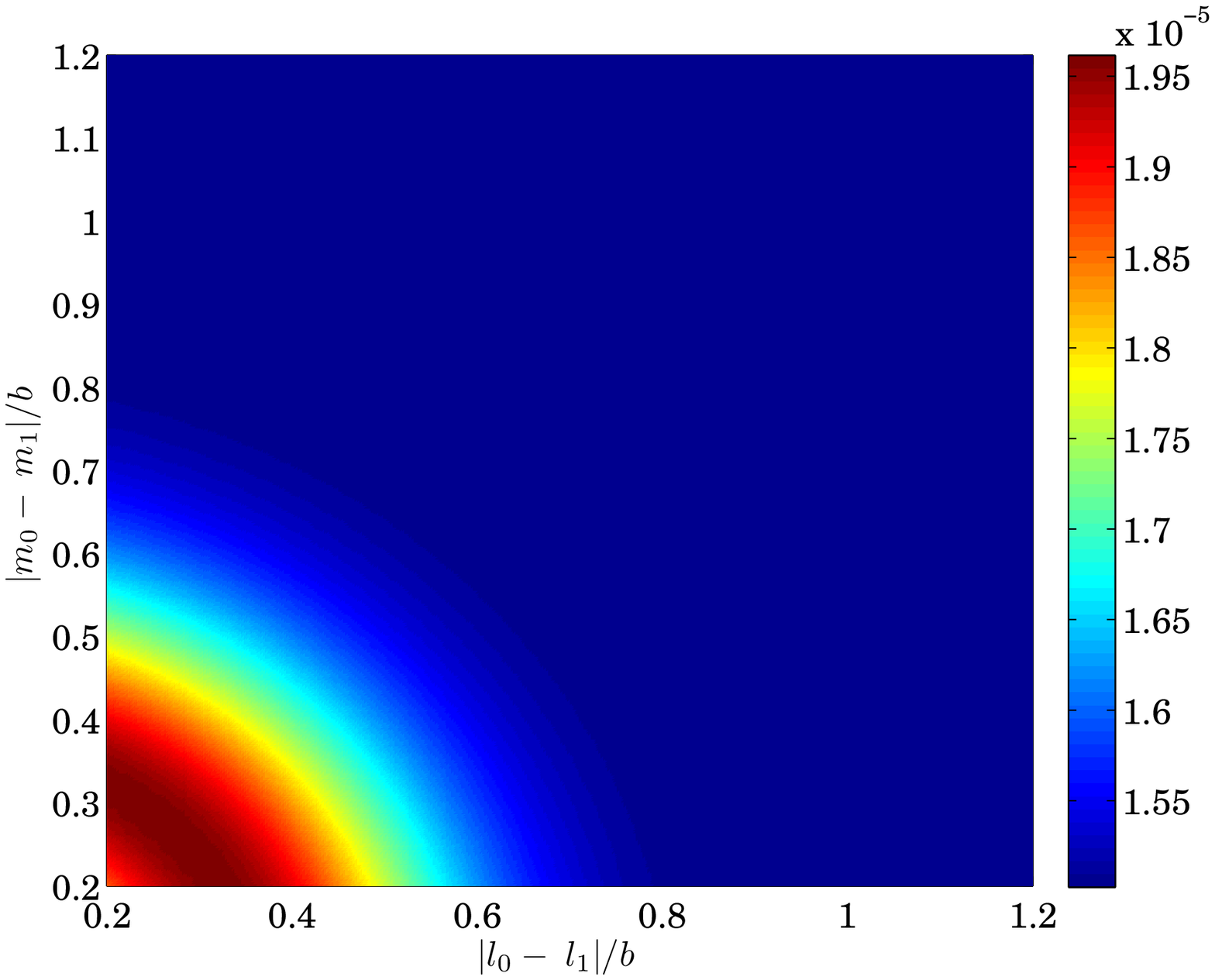,width=3.7cm}}
\vspace{0.1cm} \centerline{(b)}\smallskip
\end{minipage}
\end{minipage}
\caption{Cramer Rao lower bounds for a two dimensional case. The $uv$ coverage is given in (a). In (b), the total variance in estimating a source position, $Var(l_0)+Var(m_0)+Var(l_1)+Var(m_1)$ is given. The axes indicate the separation of the two sources, i.e. $|l_0-l_1|/b$ and $|m_0-m_1|/b$. The amplitudes are fixed at  $\gamma_0=1.0$, $\gamma_1=0.4$, while the noise $\sigma=0.2$. The nominal resolution is $b=1/max(\sqrt{u^2+v^2})$.\label{two_crlb_2d}}
\end{figure}

It is straightforward to extend the results derived in (\ref{crlb2}) for a two dimensional visibility plane, where $u$ and $v$ are the visibility axes. We again consider two sources, with magnitudes $\gamma_0$,$\gamma_1$, positioned at $(l_0,m_0)$ and $(l_1,m_1)$ respectively. The sampled visibility at the $i$-th point on the $uv$-plane is given by
\beq \label{vis2d}
y_i=\gamma_0 \exp^{\bigl(-j2\pi (l_0 u_i+m_0 v_i)\bigr)}+ \gamma_1 \exp^{\bigl(-j2\pi (l_1 u_i+m_1 v_i)\bigr)} + n_i
\eeq
As shown in \cite{SHPfull}, we can derive bounds for the parameter set ${\bmath \theta}=[l_0,m_0,l_1,m_1,\gamma_0,\gamma_1]$. We have given a numerical example in Fig. \ref{two_crlb_2d} for this case. As seen on Fig. \ref{two_crlb_2d} (b), as the two sources come closer, the variance in estimating their positions increase. 

As discussed in \cite{Briggs}, any extended source could be represented by clean components equivalent to the Fourier components of the brightness distribution of that source. However, the accuracy of this representation is limited  when we have finite resolution due to lower bounds in estimation of positions and magnitudes of those clean components. Thus, it is futile to make the image grid arbitrarily small hoping the accuracy of our modeling of extended sources improve. Any pixel based deconvolution algorithm would run into this limitation and this forces us to find alternative methods to model such sources.

\subsection{Deconvolution using an arbitrary basis}
Because of the limitation of modeling an extended source by using multiple clean components, we strive to improve this by using other forms of image representation.
Let us define an arbitrary basis as
\beq
{\mathbb S}=\{s_1(u),s_2(u),\ldots,s_K(u)\}
\eeq
where $s_k(u)$ is the $k$-th basis function at $u$, on the visibility plane (axis). If we observe an extended source, the $i$-th visibility point can be given as
\beq \label{sumu}
y_i=\sum_k \theta_k s_k(u_i)+n_i
\eeq
where ${\bmath \theta}=[\theta_1, \theta_2,\ldots, \theta_K]^T$ are the $K$ parameters we need to estimate. Representing the bases $\mathbb S$ evaluated at $u$ by ${\bf s}(u)=[s_1(u),s_2(u),\ldots,s_K(u)]^T$, we have the vectorized form
\beq
y_i={\bf s}^{T}(u_i) {\bmath \theta} + n_i
\eeq
and combining all visibility points in vector ${\bf y}$ we have
\beq \label{lst}
{\bf y}={\bf S} {\bmath \theta} +{\bf n}
\eeq
where ${\bf S}=[{\bf s}(u_0),{\bf s}(u_1),\ldots,{\bf s}(u_N)]^{T}$ and ${\bf n}\buildrel\triangle\over=[n_1,n_2,\ldots,n_N]^T$.

This is the well studied linear statistical model and the likelihood can be expressed as
\beq
p({\bf y}|{\bmath \theta},\sigma^2)=\frac{1}{\pi^N \det{(\sigma^2 {\bf I})}} \exp\biggl(\frac{-1}{\sigma^2}({\bf y}-{\bf S}{\bmath \theta})^H({\bf y}-{\bf S}{\bmath \theta})\biggr)
\eeq
The ML estimate is $\widehat{\bmath \theta} ={\bf S}^{\dagger}{\bf y}$
where ${\bf S}^{\dagger}$ is the matrix pseudo inverse of ${\bf S}$.
From \cite{Kay}, we get the Cramer-Rao lower bound as
\beq \label{Varth}
Cov({\bmath \theta}) =({\bf S}^H(\sigma^2 {\bf I})^{-1}{\bf S})^{-1} =\sigma^2 ({\bf S}^H{\bf S})^{-1}.
\eeq
Using (\ref{Varth}), we get the variance of estimation of the $k$-th parameter $\theta_k$ as the $k$-th diagonal entry of $\sigma^2 ({\bf S}^H{\bf S})^{-1}$. Note that this is minimized if ${\bf S}^H{\bf S}={\bf I}$. In other words, if we choose the basis such that ${\bf S}$ is unitary, we get the lowest error. This is the primary motivation behind having an orthonormal basis instead of clean components.

\subsection{Information theoretic bounds}
An important question we should answer is the maximum number of basis functions or clean components that can be used to represent any given source. Following the arguments presented in \cite{Slepian}, we see that most sources have compact support both in the image plane and the Fourier plane. In the latter case, the support is also limited by the distribution of sampling points (baselines). Hence we can use Landau Pollak theorem \cite{Landau} to limit the degrees of freedom of any source that can be seen from any interferometer. If the support area in the image plane is $A_{im}$ and the corresponding support in the Fourier plane is $A_{uv}$, then the number of degrees of freedom is bounded by $A_{im}\times A_{uv}$. This can be used as a criterion to limit the number of clean components (hence the pixel size) as well as the number of basis functions that can be effectively used to model any given source.
\section{Results}
\begin{figure}[htbp]
\begin{minipage}{0.98\linewidth}
\begin{minipage}{0.98\linewidth}
\centering
 \centerline{\epsfig{figure=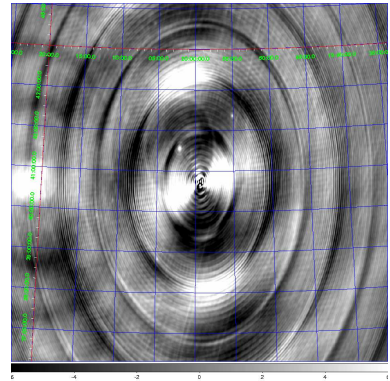,width=5.0cm}}
\vspace{0.1cm} \centerline{(a)}\smallskip
\end{minipage}
\vspace{0.1cm}
\begin{minipage}{0.98\linewidth}
\centering
 \centerline{\epsfig{figure=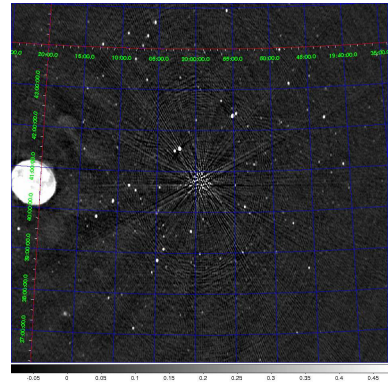,width=5.0cm}}
\vspace{0.1cm} \centerline{(b)}\smallskip
\end{minipage}
\end{minipage}
\caption{Deconvolved images of the area surrounding Cygnus A (a) using CLEAN (about 1000 clean components) and (b) using shapelet deconvolution (about 400 modes). The dynamic range of (a) is about 10,000 while in (b) it is about 500,000. Cygnus A (peak flux 10 kJy) is at the center of the image and has been subtracted. The noise in (b) is about 20 mJy. Far more fainter sources are seen on (b) compared to (a).\label{cyga_example}}
\end{figure}

As an example, we present results of an observation of Cygnus A, using the Westerbork Synthesis Radio Telescope. Around 150 MHz, the source Cygnus A is barely resolved by the WSRT. Hence, traditional clean based algorithms fail to perform satisfactorily. In Fig. \ref{cyga_example} (a), we have given the results obtained using clean. In this case, the dynamic range (ratio of the peak flux to the noise in the image) is about 10,000. However, by using shapelet basis functions we can improve the result to reach a dynamic range of well over 500,000 as seen in Fig. \ref{cyga_example} (b).

\section{Conclusions}
We have presented limitations of pixel based deconvolution of extended sources in radio interferometry. We have both theoretically and based on real data, shown that by using suitable orthonormal basis functions, we could overcome this limitation. Although we have chosen shapelets as our example basis functions, future work should focus on finding better basis functions in terms of performance and in terms of computational efficiency.

\bibliographystyle{IEEEtran}

\end{document}